\documentclass[preprint,11pt]{JHEP3}% 10pt is ignored!

\JHEPspecialurl{http://jhep.sissa.it/JOURNAL/JHEP3.tar.gz}

\usepackage{epsfig,multicol,amsmath}

\usepackage{epstopdf}
\usepackage{bm}  % for the command \bm (boldface greek letters)
\usepackage{amsmath}     % defines the command \dfrac (fractions in displaystyle)
\usepackage{epsfig,epsf}
\usepackage{graphicx}
\usepackage{rotating}
\usepackage{showkeys}
\usepackage{cite}

\def\beq{\begin{equation}}
\def\eeq{\end{equation}}
\def\bea{\begin{eqnarray}}
\def\eea{\end{eqnarray}}

\def\eq#1{{Eq.~(\ref{#1})}}
\def\fig#1{{Fig.~\ref{#1}}}

\newcommand{\Lb}{\left(}
\newcommand{\Rb}{\right)}
\parskip=\itemsep              

\newcommand{\h}{\frac{1}{2}}

\newcommand{\y}{{\cal Y}}

\def\pom{{I\!\!P}}

\title{ QCD motivated approach to soft interactions at high energies: 
nucleus-nucleus and hadron-nucleus collisions}
\author{\Large  E. Gotsman\thanks{Email:
gotsman@post.tau.ac.il.}\,,A. Kormilitzin\thanks{Email:
andrey1@post.tau.ac.il.},  E. Levin\thanks{Email:
leving@post.tau.ac.il.}\,\,and\,\,U. Maor\thanks{Email: maor@post.tau.ac.il.}\, 
\\
\,\,Department of Particle Physics, School of Physics and Astronomy\\
Raymond and Beverly Sackler Faculty of Exact Science\\  
Tel Aviv University, Tel Aviv, 69978, Israel\\}

\abstract
{ In this paper we consider
 nucleus-nucleus and hadron-nucleus reactions in the kinematic region: $g
\,A^{1/
3}\,G_{3\pom}\,\exp\Lb \Delta Y\Rb \approx 1\,\,\,\mbox{and} \,\,\,
G^2_{3\pom}\,\exp\Lb \Delta Y\Rb \approx 1 $, where $G_{3\pom}$ is the
triple
Pomeron coupling, $g$ is the vertex of Pomeron nucleon interaction, and
 1 + $\Delta_{\pom}$ denotes the Pomeron intercept. We find that in this
kinematic region
 the traditional Glauber-Gribov eikonal approach is inadequate.
We show that it is necesssary to take into account  inelastic Glauber
corrections, which can not be expressed in terms of the
nucleon-nucleon scattering amplitudes. In the wide range of energies where
$\alpha'_\pom \,Y \,\ll\,R^2_A$,
the scattering amplitude for the nucleus-nucleus interaction, does not
depend on
the details of the nucleon-nucleon interaction at high energy.
In the formalism we  present, the only (correlated) parameters that are
required
to describe the data
 are $\Delta_{\pom}$,  $G_{3\pom}$ and $g$. These parameters were taken
from our description of the nucleon-nucleon data at high energies \cite{GLMM}.
The predicted nucleus modification factor is
compared with
RHIC Au-Au data at $W\,=\,200\,GeV.$
  Estimates  for  LHC energies are presented and  discusssed.}

\dedicated{PACS: 13.85.-t, 13.85.Hd, 11.55.-m, 11.55.Bq}
\preprint{TAUP -2907-09\\
{\tt }\\
\today}
\begin{document}
%%%%%%%%%%%%%%%%%%%%%%%%%%%%%%%%%%%%%%%%%%%%%%%%%%%%%%%%
\section{Introduction}
%%%%%%%%%%%%%%%%%%%%%%%%%%%%%%%%%%%%%%%%%%%%%%%%%%%%%%%%%%%%
\par
The main goal of this paper is to generalized our approach to soft 
interactions developed in Ref.\cite{GLMM},
to nucleus-nucleus  and hadron-nucleus interactions. 
This approach is based on two main assumptions that 
provide a natural bridge to the high density QCD approach 
(see Refs.\cite{BFKL,LI,GLR,MUQI,MV,B,K,JIMWLK}).
i) $\alpha'_\pom = 0$; and ii) All Pomeron-Pomeron interactions can be 
constructed from triple Pomeron vertices through Fan diagrams. 
\par
Based on the above two assumption, we have analyzed in Ref.\cite{GLMM} 
the available data on $p-p$ and $\bar{p}-p$ soft scattering so as to determine 
the soft Pomeron features. We obtain:
\newline
1) $\Delta_{\pom}\,=\,0.35$. 
\newline
2) $\alpha'_\pom = 0.012$. This fitted value supports our input assumption.
\newline
3) The value of the triple Pomeron vertex coupling 
$G_{3\pom}\,=\,\gamma \Delta_{\pom}$ is small (the fitted $\gamma\,=\,0.0242$).
\newline
4) Note that Pomeron-hadron (and Regge-hadron) interactions are treated in this 
approach phenomenologically.
\newline
To summarize: The data analysis\cite{GLMM} confirms our input assumptions   
and leads to a natural matching between the soft Pomeron and the pQCD 
hard Pomeron. Indeed $\Delta_{\pom}\,\approx \,\alpha_S$, 
$\gamma\,\approx\,\alpha_S^2$ and $\alpha'_{\pom}\,\ll\,1$. 
\par
In section 2 we derive the main equations governing nucleus-nucleus  
scattering at high energy. To this end we define the kinematic regions 
in which our formalism is applicable, 
\beq \label{I1}
g \,A^{1/3}\,G_{3\pom}\,\exp\Lb \Delta Y\Rb \approx 1\,\,\,\mbox{and} \,\,\,
G^2_{3\pom}\,\exp\Lb \Delta Y\Rb \approx 1.
\eeq
$1\,+\,\Delta_{\pom}$ denotes the intercept of the soft Pomeron, 
$g$ is the vertex coupling of the Pomeron-nucleon interaction and  
$G_{3\pom}$ is the vertex coupling of the triple Pomeron interaction. 
As we shall see, the kinematic region defined by \eq{I1} is wider than 
the kinematic region relevant to hadron-hadron scattering. Consequently, 
we have to go beyond the traditional Glauber-Gribov eikonal approach.
The result we obtain suggests that it is necessary
 to take into account the inelastic Glauber
correction which can not be expressed in terms  of the
nucleon-nucleon scattering amplitudes. In the wide range of energies where
$\alpha'_\pom \,Y \,\ll\,R^2_A$,
the scattering amplitude for the nucleus-nucleus interaction does not depend on
the details of the nucleon-nucleon interaction at high energy.
In the formalism we  present, the only (correlated) parameters that we need
to know are $\Delta_{\pom}$, $\alpha'_\pom$, $G_{3\pom}$ and $g$.
 $\Delta_{\pom}$, $G_{3\pom}$ and $g$ were obtained from the
data analysis of proton-proton soft scattering\cite{GLMM}. Our basic dynamical
assumption is that $\alpha'_\pom\,=\,0$, which is supported by the fitted value
of $\alpha'_\pom\,=\,0.012$ obtained in Ref.\cite{GLMM}.
Since the fitted values of $G_{3\pom}$ and $\alpha'_\pom$
are small, \eq{I1} is valid over a  wide range of energies,
including the LHC energy. For the sake of completeness, we discuss in
section 3 the main equations for hadron-nucleus interactions
that have been derived in the kinematic region of \eq{I1}
in Refs.\cite{SCHW,BGLM}.
In section 4 we adjust the general formulae of sections 3 and 4
to the specific approach of Ref.\cite{GLMM}. 
Section 5 is devoted to a comparison of our results 
with the experimental data, mostly on the nuclear modification
factor in the RHIC range of energies. Predictions for LHC energies 
are presented and discussed.
In the conclusions we reflect on the physical meaning of our 
approach and its relation to the Color
Glass Condensate (CGC) model\cite{MV}.
%%%%%%%%%%%%%%%%%%%%%%%%%%%%%%%%%%%%%%%%%%%%%%%%%%%%%%%%%%%%
\section{Equations for nucleus-nucleus collisions}
%%%%%%%%%%%%%%%%%%%%%%%%%%%%%%%%%%%%%%%%%%%%%%%%%%%%%%%%%%%%
\par
In the framework of the Pomeron Calculus \cite{GRIBRT} 
(see also Refs.\cite{COL,SOFT,LEREG})
there are two different kinematic domains whereone  can develop 
a theoretical approach for nucleus-nucleus scattering. 
\par
In the first domain  we  consider 
\beq \label{KR1}
g_1\,g_2\,\int d^2 b' \,d^2 b"\, S_{A_1}(b'')\,S_{A_2}( \vec{b} - 
\vec{b}')\,P(Y,\vec{b}" - \vec{b}')\,=\,
g_1\,g_2 \,\int \,d^2 b"\,P(Y,b")\,\int d^2 b' \,\, S_{A_1}(b')\,S_{A_2}
(\vec{b} - \vec{b}')\,\,
\eeq
$$\propto\,\,\,\,
g_1\,g_2 \,A^{1/3}_1\,A^{1/3}_2\,\Lb R^2_{A_1} \,+\,R^2_{A_2}\Rb\,
e^{\Delta_{\pom} Y}\,\,\,
\approx\,\,\,\,1;
$$
$$
g_{i}\,S_{A_i}(b)\,G_{3\pom}\,e^{\Delta_{\pom} Y} \,\,
\propto\,\,g_i\,G_{3\pom} A^{1/3}_i\,e^{\Delta_{\pom} Y} 
\,\,\ll\,\,1;\,\,\,\,\,\,\,\,
G^2_{3\pom}\,e^{\Delta_{\pom} Y} \,\,\ll\,\,1. 
$$
In this kinematic region the main contribution 
stems from the diagrams of \fig{glset}. Summing these diagrams, 
we obtain the Glauber-Gribov eikonal expressions describing  
nucleus-nucleus scattering\cite{GLAUB,GRIBA}. In this approach the 
nucleus-nucleus amplitude can be derived from the knowledge of the 
nucleon-nucleon cross section. Specifically, we get
\bea \label{GG}
N_{el}\Lb Y; b\Rb\,\,&=&\,\,i \Lb 1 \,\,-\,\,
\exp\Lb -\h g_1\,g_2\,\int d^2 b'\,
S_{A_1}\Lb b'\Rb\,S_{A_2}\Lb \vec{b} - \vec{b}'\Rb\,
\int d^2 b" \,P\Lb Y,b"\Rb\Rb\Rb\,\,\nonumber\\
&=&\,\,
\Lb 1 \,\,-\,\,\exp\Lb -\h\, g_1\,g_2\,
\int d^2 b'\,S_{A_1}\Lb b'\Rb\,S_{A_2}\Lb \vec{b} - \vec{b}'\Rb\,\,
e^{\Delta Y}\Rb\Rb.
\eea
\par
In the second kinematic region 
\beq \label{KR2}
g_1\,g_2\,\int d^2 b' \,d^2 b"\, S_{A_1}(b'')\,S_{A_2}
( \vec{b} - \vec{b}')\,P(Y,\vec{b}" - \vec{b}')\,=\,
g_1\,g_2 \,\int \,d^2 b"\,P(Y,b")\,\,\int d^2 b' \,\, S_{A_1}(b')\,S_{A_2}
( \vec{b} - \vec{b}')\eeq
$$\,\,\propto\,\,\,\,
g_1\,g_2 \,A^{1/3}_1\,A^{1/3}_2\,\Lb R^2_{A_1} \,
+\,R^2_{A_2}\Rb\,e^{\Delta_{\pom} Y}\,\,\,>\,\,\,1;
$$
$$
g_{i}\,S_{A_i}(b)\,G_{3\pom}\,e^{\Delta_{\pom} Y} \,\,
\propto\,\,g_i\,G_{3\pom} A^{1/3}_i\,e^{\Delta_{\pom} Y} \,\,
\approx \,\,1;\,\,\,\,\,\,\,\,
G^2_{3\pom}\,e^{\Delta_{\pom} Y} \,\,\ll\,\,1,
$$
it is necessary to consider the the more complicated 
set of diagrams presented in \fig{netset}.
We conclude that the knowledge of the nucleon-nucleon amplitude 
is not sufficient to enable a calculation of nucleus-nucleus scattering. 
In addition to the above, we also need to know the structure of the 
Pomeron-Pomeron interactions, so as to tackle 
the problem of the summation of these diagrams. 
In \fig{netset} we assume that only the triple
Pomeron vertex contributes to multi Pomeron interactions.
%%%%%%%%%%%%%%%%%%%%%%%%%%%%%%%%%%%%%%%%%%%%%%%%%%%%%%%%%%%%
\FIGURE[ht]{
\centerline{\epsfig{file=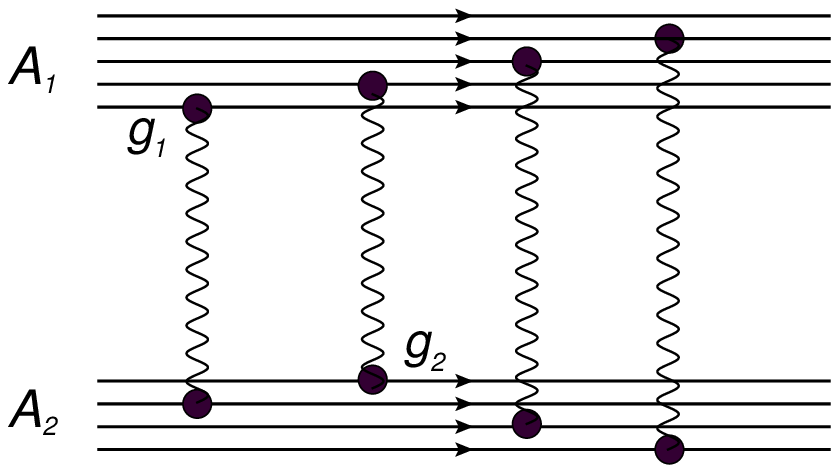,width=100mm}}
\caption{The full set of the diagrams for Glauber-Gribov approach 
which contribute to the scattering amplitude in the kinematic region of \eq{KR1}}
\label{glset}
}
%%%%%%%%%%%%%%%%%%%%%%%%%%%%%%%%%%%%%%%%%%%%%%%%%%%%%%%%%%%%
%%%%%%%%%%%%%%%%%%%%%%%%%%%%%%%%%%%%%%%%%%%%%%%%%%%%%%%%%%%%
\FIGURE[ht]{
\centerline{\epsfig{file=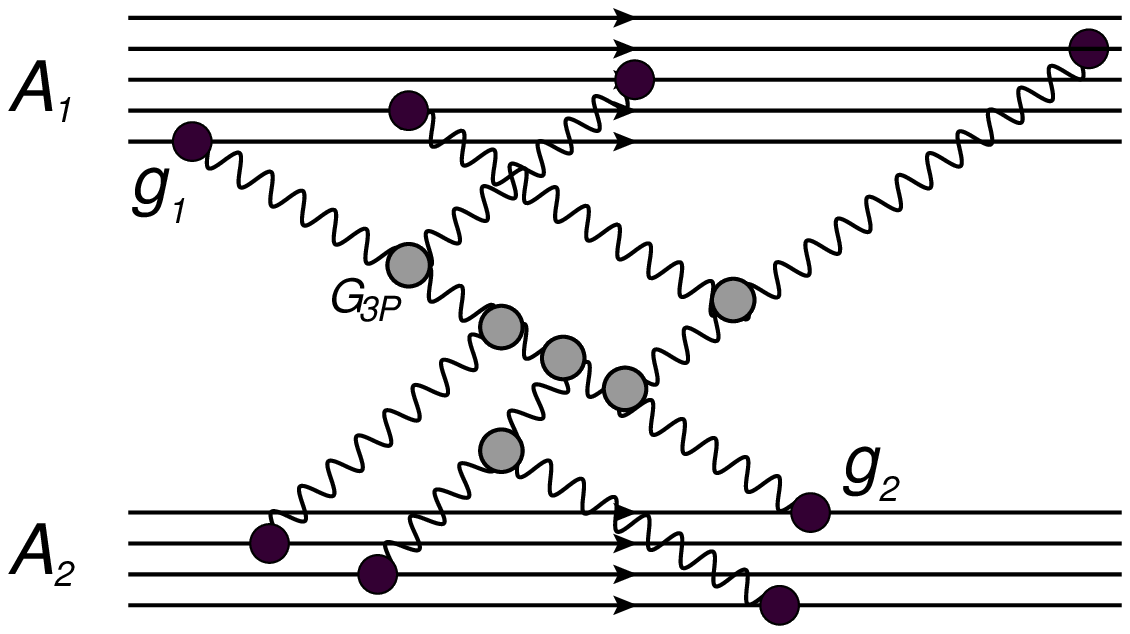,width=100mm}}
\caption{The full set of the diagrams that contribute to the 
scattering amplitude for the kinematic
region \eq{KR2}.}
\label{netset}
}
%%%%%%%%%%%%%%%%%%%%%%%%%%%%%%%%%%%%%%%%%%%%%%%%%%%%%%%%%%%%
\par
As stated, the main goal of this paper is to calculate 
the nucleus-nucleus amplitude in the 
kinematic region of \eq{KR2}.
We start with the MPSI approach \cite{MPSI}.
%%%%%%%%%%%%%%%%%%%%%%%%%%%%%%%%%%%%%%%%%%%%%%%%%%%%%%%%%%%%
\subsection{MPSI approach}
%%%%%%%%%%%%%%%%%%%%%%%%%%%%%%%%%%%%%%%%%%%%%%%%%%%%%%%%%%%%
The MPSI approach is based on the observation 
that for a Pomeron with a intercept larger than 1, 
i.e. $\Delta_{\pom} > 0$, the main diagrams of interest 
have the form of \fig{mpsiset} with an arbitrary $y$ which 
is of the order of $Y$. We have discussed this approach 
in our previous publication\cite{GLMM,LMP}. 
The method of calculating the sum of \fig{mpsiset} diagrams  
is based on $t$-channel unitarity 
adjusted \cite{MPSI} to the summation of Pomeron diagrams.
\par
The general formula has the form
\bea \label{MPSI}
N^{MPSI}_{el} \left(Y\right)\,\,\,&=&
\,\,\sum^{\infty}_{n=1}\,\,(-1)^{n + 1}\frac{1}{n!}\,
\,\,\gamma^n\,\,  \frac{\partial^n\,N^{MFA}(Y -y,\,\gamma^p_R)}{\partial^n\, 
\gamma^p_R}|_{\gamma^p_R =0}
\,\, \frac{\partial^n\,N^{MFA}(y,\,\gamma^t_R)}{\partial^n\, 
\gamma^t_R}|_{\gamma^t_R =0}  \\
 &=& \,\,1\,\,\,-\,\,\,\left\{\exp \Lb
-\,\gamma\,\frac{\partial}{\partial \,\gamma^p_{R}}\,
\frac{\partial}{\partial \,\gamma^t_{R }}\Rb\,\,\,
N^{MFA}\Lb Y  - y,\, \gamma^p_{R}\Rb\,\,N^{MFA}\Lb y,  \gamma^t_{R}\Rb
\,\right\}\mid_{\gamma^p_{R} =0; \,\gamma^t_{R}  = 0}. \notag
\eea
$\gamma^{p,t}_R$ are related to projectile and target, 
respectively, and  $\gamma$ is the
amplitude for 'wee' partons (colorless dipoles in QCD) 
scattering at low energies. 
$N^{MFA} $ denotes the amplitude, which for a nucleus has the form 
\beq \label{MPSI1}
N^{MFA}_{A_i}\Lb y, b \Rb \,\,=\,\,1 \,-\,\exp\Lb - \tilde{g}_i\,
S_{A_i}\Lb b \Rb\,N^{MFA}_N\Rb.
\eeq
Here, $N^{MFA}_N$ denotes the sum of \fig{fanset} diagrams  
where $\tilde{g}_1\tilde{g}_2 \gamma = g_1 g_2$.
In the case of $\alpha'_\pom\, \ll\, 1$, 
i.e. for energies $\alpha'_\pom Y < R^2_A$, 
this sum is 
\beq \label{MPSI2}
N^{MFA}_N\,\,=\,\,\frac{ \gamma_R\,e^{\Delta_{\pom} (Y - y)}}{ 1\,\,+\,\, 
\gamma_R\,e^{\Delta_{\pom} (Y - y)}},
\eeq
where $\gamma_R$ denotes the scattering amplitude at low energy of 
the `wee' parton (colorless dipole in QCD) with the target. 
%%%%%%%%%%%%%%%%%%%%%%%%%%%%%%%%%%%%%%%%%%%%%%%%%%%%%%%%%%%%%%%%%%%%%%%%
\DOUBLEFIGURE[ht]{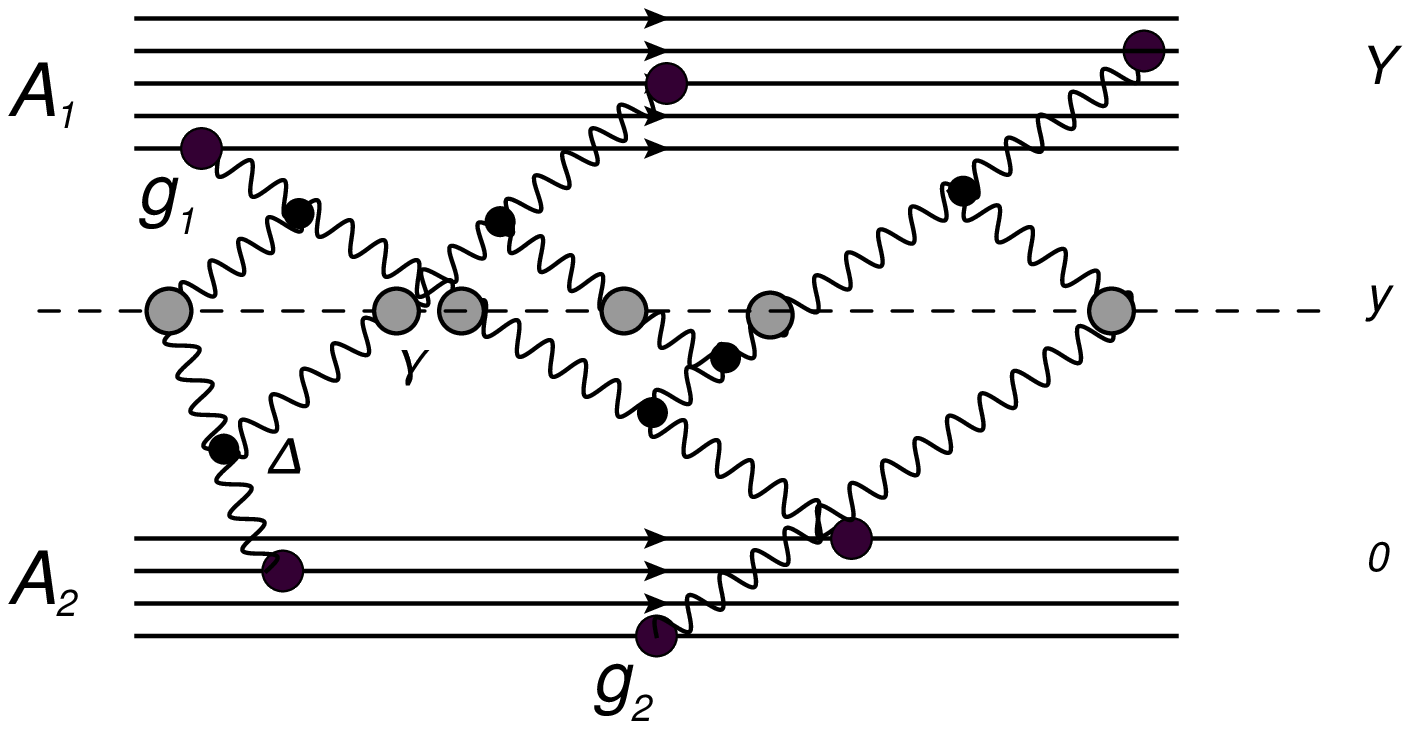,width=95mm}{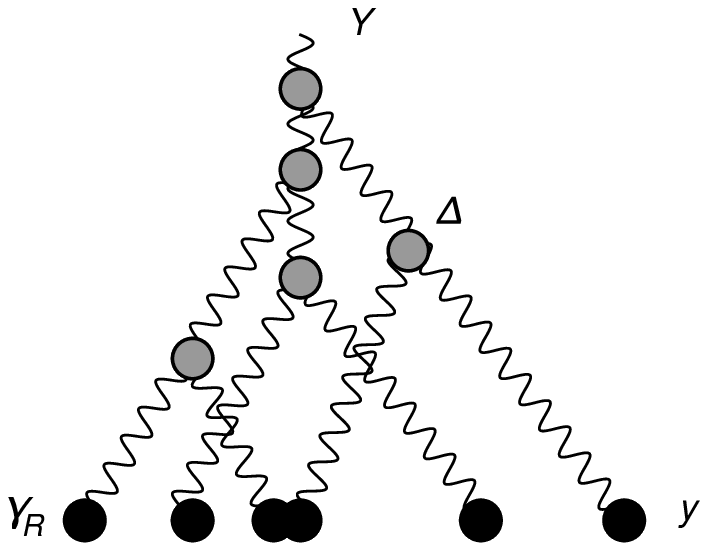,width=65mm,height=45mm}
{The set of diagrams that contribute to the scattering amplitude 
in the MPSI approximation for the kinematic region \eq{KR2}.\label{mpsiset}}
{The set of fan diagrams that contribute to $N^{MFA}_N$\label{fanset}}
%%%%%%%%%%%%%%%%%%%%%%%%%%%%%%%%%%%%%%%%%%%%%%%%%%%%%%%%%%%%%%%%%%%%%%%%%
\par
Using a generating function for Laguerre polynomials (see Ref.\cite{RY} formula 
{\bf 8.973(1)}), 
\beq \label{MPSI3}
(1 - z)^{- \alpha - 1}\, \exp\Lb \frac{x\,z}{z - 1}\Rb\,\,\,=
\,\,\,\sum^{\infty}_{n = 0}\,L^{\alpha}_n\Lb x \Rb \,z^n,
\eeq
we obtain for \eq{MPSI2}
\beq \label{MPSI4}
N^{MFA}\Lb Y-y; \gamma_R\Rb\,\,\,=\,\,\,-\,\sum^\infty_{n=1}\,
L^{-1}_n\Lb\tilde{g}_i\Rb\,\Lb -  \gamma_R e^{\Delta_{\pom} (Y - y)}\Rb^n. 
\eeq
For simplicity we have omitted the $b$ dependence which is 
easy to include, as each 
$\tilde{g}_i$ should be multiplied by $S_{A_i}(b)$.
\par
Using \eq{MPSI3} and \eq{MPSI} we have for the scattering amplitude,
\beq \label{MPSI5}
N^{MFA}\Lb Y \Rb\,\,=\,\,\sum^{\infty}_{n=0}\,n!\,
L^{-1}_n\Lb\tilde{g}_i\Rb\,\,L^{-1}_n\Lb\tilde{g}_k\Rb
\Lb - \gamma \,e^{\Delta_{\pom} Y}\Rb^n.
\eeq
Introducing $n! = \int^\infty_0 \,\xi^n\,e^{-\xi} \,d \xi$,  
we can re-write \eq{MPSI5} in the form
\beq \label{MPSI6}
N^{MFA}\Lb Y; \Rb\,\,=\,\,\int^\infty_0 \,d \xi\,\,e^{-\xi} \,
d\sum^{\infty}_{n=0}\,\,L^{-1}_n\Lb\tilde{g}_i\Rb\,\,L^{-1}_n
\Lb\tilde{g}_k\Rb\,\Lb - \xi \gamma_0\,e^{\Delta_{\pom} Y}\Rb^n. 
\eeq
$\gamma_0\,=\,G_{3\pom}/\Delta_{\pom}$ is obtained from the data analysis of 
of soft proton-proton scattering\cite{GLMM}.
Using formula {\bf 8.976(1)} of Ref.\cite{RY},
\beq \label{SUML}
\sum^\infty_{n = 0}\,n!\,z^n\,\frac{L^{\alpha}_n(x)\,L^{\alpha}_n(y)}
{\Gamma\Lb n + \alpha +1\Rb}\,\,\,=\,\,\frac{\Lb x\,y\,z\Rb^{- \frac{1}{2}\alpha}}
{1 - z}\,\exp\Lb - z \frac{x + y}{1 - z}\Rb\,I_\alpha\Lb 2 \frac{\sqrt{x\,y\,z}}
{1 - z }\Rb, 
\eeq
we derive the final result
\beq \label{MPSI7}
N^{MFA}_{i,k}\Lb Y \Rb\,\,=\,\,\int^\infty_0 \,\frac{d \xi}{\xi}\,\,
e^{-\xi}\,\frac{\Lb \tilde{g}_i\,\tilde{g}_k\,\xi\,T(Y)
\Rb^{\frac{1}{2}}\,}{ 1\,+\,\xi T(Y)}\,\exp\left\{ - \xi \,T(Y)\,
\frac{\tilde{g}_i + \tilde{g}_k}{1\,+\,\xi\,T(Y)}\right\}\,J_1\Lb 
2\frac{ \sqrt{\tilde{g}_i\,\tilde{g}_i\,\,\xi\,T\Lb Y \Rb}}{1\,+\,\xi\,T(Y)}\Rb,
\eeq
where
\beq \label{T}
T\Lb Y\Rb\,\,=\,\,\gamma\,e^{\Delta_{\pom} Y}.
\eeq
Recalling that the function should depend on $\tilde{q}_1 \tilde{g}_2 \gamma$,
on $\tilde{q}_1 \gamma$ or on $\tilde{q}_2 \gamma$, one can see that
\eq{MPSI7} reduces to a simple and elegant formula for 
the case where we have $\tilde{g}_i\,T(Y)\,\sim \,1$, 
$\tilde{g}_i\tilde{g}_i\,T(Y)\,>\, \,1$ and $T(Y) \ll 1$. 
Indeed, after integrating over $\xi$ we get  
\beq \label{MPSI8}
A_{i,k}\Lb Y; b\Rb\,\,\,= \,\,1 \,\,\,-\,\,\exp\left\{ - 
\,\,\,\frac{\tilde{g}_i\tilde{g}_k\,T(Y)}{ 1\,+\,T(Y)\,
\left[\tilde{g}_i\,\, + \,\,\tilde{g}_k\right]}\right\}. 
\eeq
\par
For \eq{MPSI8} it is easy to write the expression that takes into account 
the correct impact parameter behavior. It has the form 
\beq \label{MPSI91}
A\Lb Y; b\Rb\,\,\,= \,\,i\Lb 1 \,\,\,-\,\,\exp\left\{ - \h\, \int d^2 b'\,
\,\,\,\frac{\Lb \tilde{g}_1 S_{A_1}\Lb\vec{b}'\Rb\,\tilde{g}_2\,S_{A_2}\Lb\vec{b} - 
\vec{b}'\Rb\,T(Y)\Rb}{ 1\,+\,T(Y)\,\left[\tilde{g}_1\,S_{A_1}\Lb\vec{b}'
\Rb + \tilde{g}_2\,S_{A_2}\Lb\vec{b} - \vec{b}'\Rb\right]}\right\}\Rb.
\eeq
%%%%%%%%%%%%%%%%%%%%%%%%%%%%%%%%%%%%%%%%%%%%%%%%%%%%%%%%%%%%
\FIGURE[ht]{
\centerline{\epsfig{file=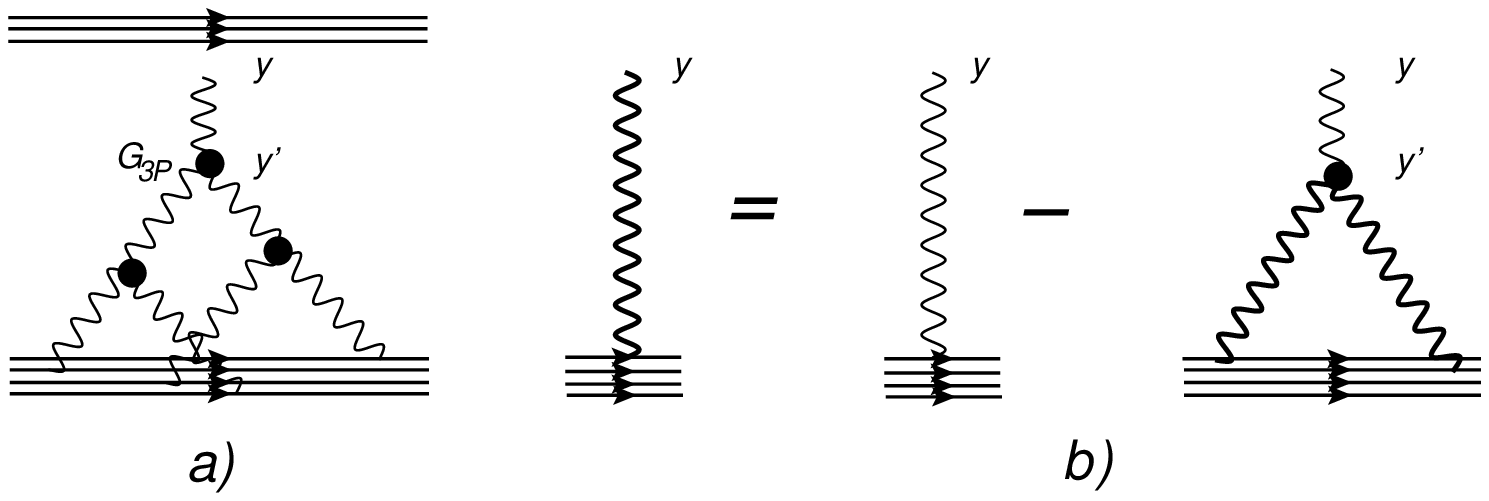,width=160mm}}
\caption{The diagram set that contributes to the scattering 
amplitude in the  approximation
in which only a merging of two Pomeron into a single Pomeron has been taken 
into account in the kinematic region of \eq{KR2}.}
\label{mpsieqset}}
%%%%%%%%%%%%%%%%%%%%%%%%%%%%%%%%%%%%%%%%%%%%%%%%%%%%%%%%%
%%%%%%%%%%%%%%%%%%%%%%%%%%%%%%%%%%%%%%%%%%%%%%%%%%%%%%%%%
\FIGURE[ht]{
\centerline{\epsfig{file=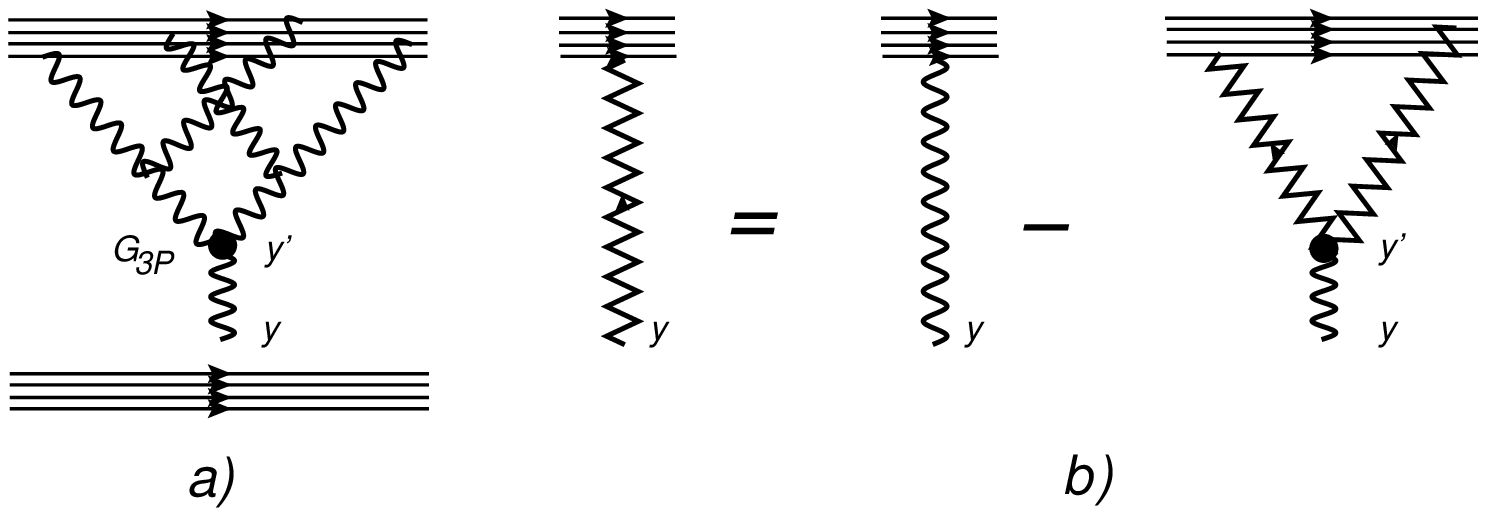,width=160mm}}
\caption{The  set of the diagrams that contributes to the 
scattering amplitude in the  approximation in which both   
the decay of a single Pomerons into two Pomerons, 
and the merging of two Pomerons into a single 
Pomeron have been taken into account
in the kinematic region of \eq{KR2}.}
\label{fulleqset}
}
%%%%%%%%%%%%%%%%%%%%%%%%%%%%%%%%%%%%%%%%%%%%%%%%%%%%%%%%%%%%
%%%%%%%%%%%%%%%%%%%%%%%%%%%%%%%%%%%%%%%%%%%%%%%%%%%%%%%%%%%%
\FIGURE[ht]{
\centerline{\epsfig{file=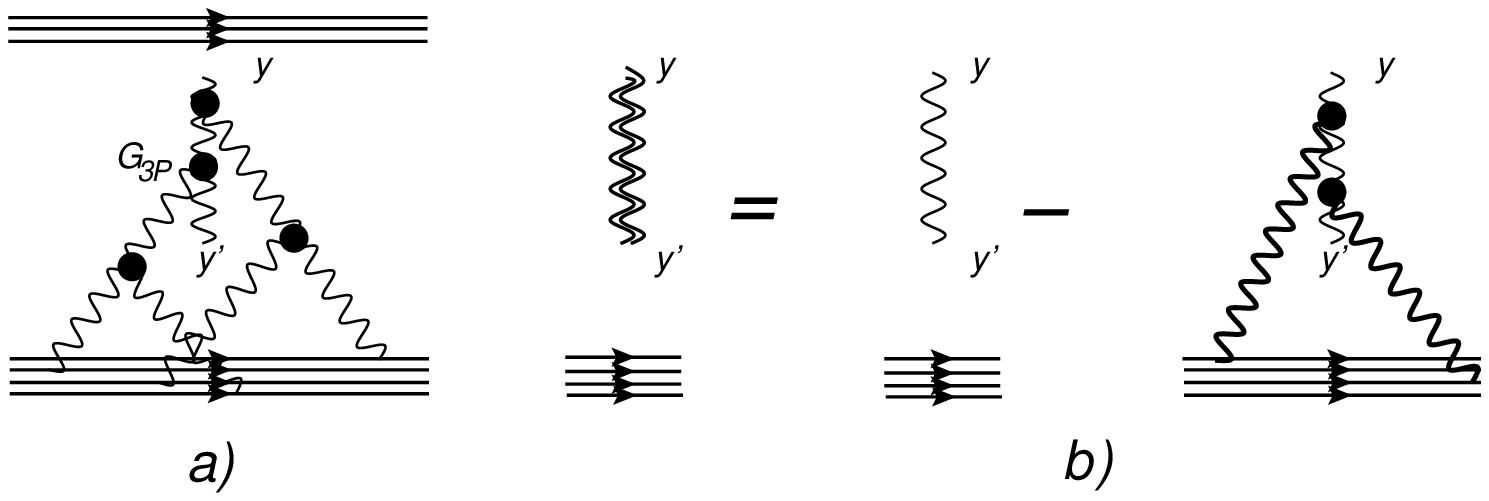,width=160mm}}
\caption{The set of diagrams that contribute to the Greem function 
$G\Lb y,y'\Rb$ in the approximation
in which only the merging of two Pomerons into a single Pomeron 
has been taken into account in 
the kinematic region of \eq{KR2}.}
\label{mpsieq1set}
}
%%%%%%%%%%%%%%%%%%%%%%%%%%%%%%%%%%%%%%%%%%%%%%%%%%%%%%%%%%%
%%%%%%%%%%%%%%%%%%%%%%%%%%%%%%%%%%%%%%%%%%%%%%%%%%%%%%%%%%%%%%}
\subsection{The complete set of equations}
%%%%%%%%%%%%%%%%%%%%%%%%%%%%%%%%%%%%%%%%%%%%%%%%%%%%%%%%%%%%%%
\par
We start  the derivation of the complete set of equations 
beyond MPSI approximation assuming that the merging of two Pomeron 
into one Pomeron is equal to zero. In this case
we need to sum the `fan' diagrams of \fig{mpsieqset}-a which we can 
do using the following equation (see \fig{mpsieqset}-b)
\beq \label{CS1}
G_{MFA}\Lb y\Rb\,\,\,=\,\,\,g_2\,e^{\Delta_{\pom} y}\,\,-\,\,G_{3\pom}\,\int^y\,d 
y'\,
e^{\Delta_{\pom} (y - y')}\,G^2_{MFA}\Lb y'\Rb,
\eeq
which can be rewritten in a differential form
\beq \label{CS2}
\frac{d G_{MFA}\Lb y\Rb}{ d y}\,\,=\,\,\Delta_{\pom} \,G_{MFA}\Lb y\Rb \,\,-\,\,
G_{3\pom}\,G^2_{MFA}\Lb y'\Rb, 
\eeq
with the initial condition 
\beq \label{CS3}
G_{MFA}\Lb y=0\Rb\,\,\,=\,\,g_2\,S_{A_2}\Lb \vec{b} - \vec{b}'\Rb.
\eeq
The solution to this equation is
\beq \label{CS4}
G_{MFA}\Lb y\Rb\,\,=\,\,\frac{g_2 \,e^{\Delta_{\pom} y}}
{ 1\, + \,g_2\frac{G_{3\pom}}{\Delta_{\pom}}\Lb 
e^{\Delta_{\pom} y}\,\,-\,\,1\Rb}. 
\eeq
We denote this solution  by a bold wave lines in 
\fig{mpsieqset},
\fig{fulleqset} and 
\fig{mpsieq1set}.
\par
We need to sum the diagrams of \fig{fulleqset}-a to obtain 
the exact two nuclei irreducible amplitude. We recall that the 
Glauber-Gribov formula (see \fig{glset}) sums all two nuclei 
reducible diagrams given by the exchange of a single Pomeron.
The equation for the exact amplitude is illustrated in graphic 
form in \fig{fulleqset}-b and has the form
\beq \label{CS5}
G_{exact}\Lb y\Rb\,\,=\,\,G_{MFA}(Y, y)\,\,-\,\,G_{3\pom}\,\int^Y_y\,d y'\,
G^2_{exact}\Lb y'\Rb\,G_{MFA}(y', y).
\eeq
The diagrams for $G_{MFA}(y', y)$ are shown in \fig{mpsieq1set}-a 
and they can be summed using the equation shown
in \fig{mpsieq1set}-b. However, for the simple case of a 
Pomeron with $\alpha'_\pom=0$ we can use the  
property of the propagator for $G_{MFA}(y)$  
\beq \label{CS6}
G_{MFA}(y', y)\,G_{MFA}(y)\,\,\,=\,\,\,G_{MFA}(y'), 
\eeq
which leads to
\beq \label{CS7}
G_{MFA}(y', y)\,\,\,=\,\,G_{MFA}(y')/G_{MFA}(y).
\eeq
Using this solution we can rewrite \eq{CS5} in the form
\beq \label{CS8}
G_{exact}\Lb y\Rb\,\,=\,\,   \frac{G_{MFA}(Y)}{G_{MFA}(y)}\,\,-\,\,G_{3\pom}\,
\frac{1}{G_{MFA}(y)}\int^Y_y\,d y'\,
G^2_{exact}\Lb y'\Rb\,G_{MFA}(y'),
\eeq
which can be written in a differential form as
\beq \label{CS9}
\frac{d G_{exact}\Lb y\Rb}{ d y}\,\,=\,\,\frac{d 
\ln G_{MFA}(y)}{d y}\, G_{exact}\Lb y\Rb\,\,\,-\,\,\,G_{3\pom}
\,G^2_{exact}\Lb y\Rb. 
\eeq
The corresponding initial condition is 
\beq \label{CS10}
 G_{exact}\Lb y=Y\Rb\,\,=\,\,g_1\,S_{A_1}\Lb b'\Rb.
\eeq
Using the solution to \eq{CS9} we can write the scattering 
amplitude which will differ from\eq{GG} by the replacement 
$P(Y) \to G_{exact}\Lb Y, b',\vec{b} - \vec{b}'\Rb$. 
It has the form
\beq \label{GGE}
 N_{el}\Lb Y; b\Rb\,\,=\,\,i \Lb 1 \,\,-\,\,\exp\Lb - \h \,
\int d^2 b'\,G_{exact}\Lb Y; b', \vec{b} - \vec{b}'\Rb\Rb\Rb.
\eeq
%%%%%%%%%%%%%%%%%%%%%%%%%%%%%%%%%%%%%%%%%%%%%% 
\subsection{MPSI solution}
%%%%%%%%%%%%%%%%%%%%%%%%%%%%%%%%%%%%%%%%%%%%%%
\par
We have not found the general solution to \eq{CS9}, 
but it is easy to demonstrate that within the MPSI approximation
this equation leads to the scattering amplitude of \eq{MPSI8}. 
First, we notice that the set of 'fan' diagrams in the MPSI approximation 
(see \fig{mpsisol})
is different from \eq{CS1} since, starting from rapidity $y"$ the 
Pomerons cannot split into two Pomerons. Therefore, the number of Pomerons 
at $y=y"$ is the same as at $y=0$. Having this in mind, we obtain 
 a solution for $G_{MFA}\Lb Y,y" \Rb$,  
\beq \label{SOLMPSI1}
G_{MFA}\Lb
Y,y"\Rb \,\,\,=\,\,\,\frac{g_1\,e^{\Delta_{\pom}\Lb Y - y"\Rb}}{1 + 
\frac{G_{3\pom}\,g_2}{\Delta_{\pom}}\Lb e^{\Delta_{\pom} Y} - 1\Rb}.
\eeq
%%%%%%%%%%%%%%%%%%%%%%%%%%%%%%%%%%%%%%%%%%%%%%%%%%%%%%%%%%%%%%%%%%%%%%
\FIGURE[h]{\begin{minipage}{80mm}{
\centerline{\epsfig{file=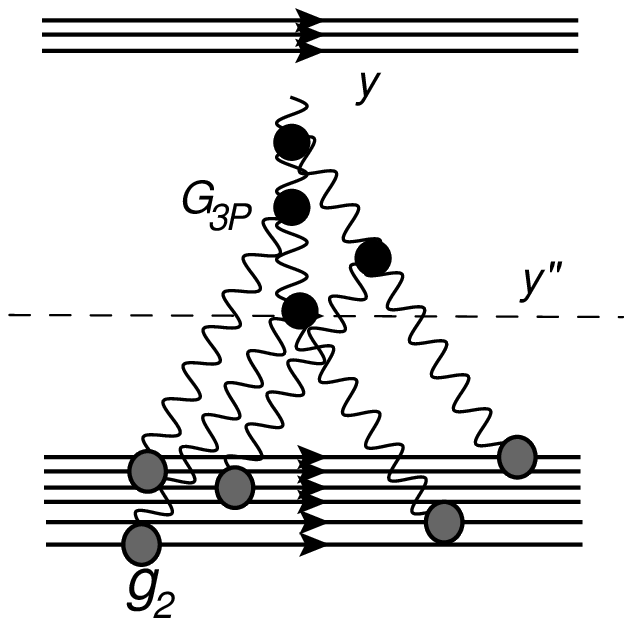,width=70mm}}
\caption{The  set of the diagrams that contribute to $G_{MFA}$ in the  
MPSI approximation.}
\label{mpsisol}}
\end{minipage}
}
%%%%%%%%%%%%%%%%%%%%%%%%%%%%%%%%%%%%%%%%%%%%%%%%%%%%%%%%%%%%%%%%%%%%%
\par
One can check that this equation sums the diagrams of \fig{mpsisol}
by expanding \eq{SOLMPSI1} with respect to 
$\Lb\frac{G_{3\pom}\,g_2}{\Delta_{\pom}}\,e^{\Delta_{\pom} Y} \Rb^n$. 
On the other hand, we can use \eq{MPSI} substituting 
$N^{MFA}(Y - y",\gamma^p_R)$ given by \eq{MPSI2} and taking 
$N^{MFA}( y",\gamma^t_R)\,=\,\,1 - \exp\Lb -\gamma^t_R\,e^{\Delta_{\pom} y"}\Rb.$
The second observation is that \eq{CS9} degenerates to
\beq \label{SOLMPSI2}
\frac{d G_{exact}\Lb y\Rb}{ d y}\,\,=
\,\,\Delta_{\pom} \, G_{exact}\Lb y\Rb\,\,-\,\,G_{3\pom}\,G^2_{exact}\Lb y\Rb,
\eeq
with the initial condition
\beq \label{SOLMPSI3}
G_{exact}\Lb y=Y\Rb\,\,=\,\,G_{MFA}\Lb Y,y"\Rb.
\eeq
\par
One can see that such a solution has the form
\beq \label{SOLMPSI4}
G_{exact}\Lb Y\Rb\,\,\,=
\,\,\frac{ g_1 g_2 e^{\Delta_{\pom} Y}}
{1 + \frac{G_{3\pom}}{\Delta_{\pom}}\,\Lb g_1 + g_2\Rb \,e^{\Delta_{\pom} Y}},
\eeq
which coincides with \eq{MPSI8} since 
$g_1\,g_2 \,e^{\Delta Y}\,=\tilde{g}_1\,\tilde{g}_2\,T\Lb Y\Rb$ and 
$ (G_{3\pom}/\Delta_{\pom} (g_1 + g_2)\,\,=\,\,(\tilde{g}_1\,+
\,\tilde{g}_2) \,T\Lb Y\Rb$. 
Substituting \eq{SOLMPSI4} into \eq{GGE} we obtain that \eq{MPSI91} 
presents the scattering amplitude.
%%%%%%%%%%%%%%%%%%%%%%%%%%%%%%%%%%%%%%%%%%%%%%%%%%%%
\section{Equations for hadron-nucleus collisions}
%%%%%%%%%%%%%%%%%%%%%%%%%%%%%%%%%%%%%%%%%%%%%%%%%%%
\FIGURE[h]{
\centerline{\epsfig{file=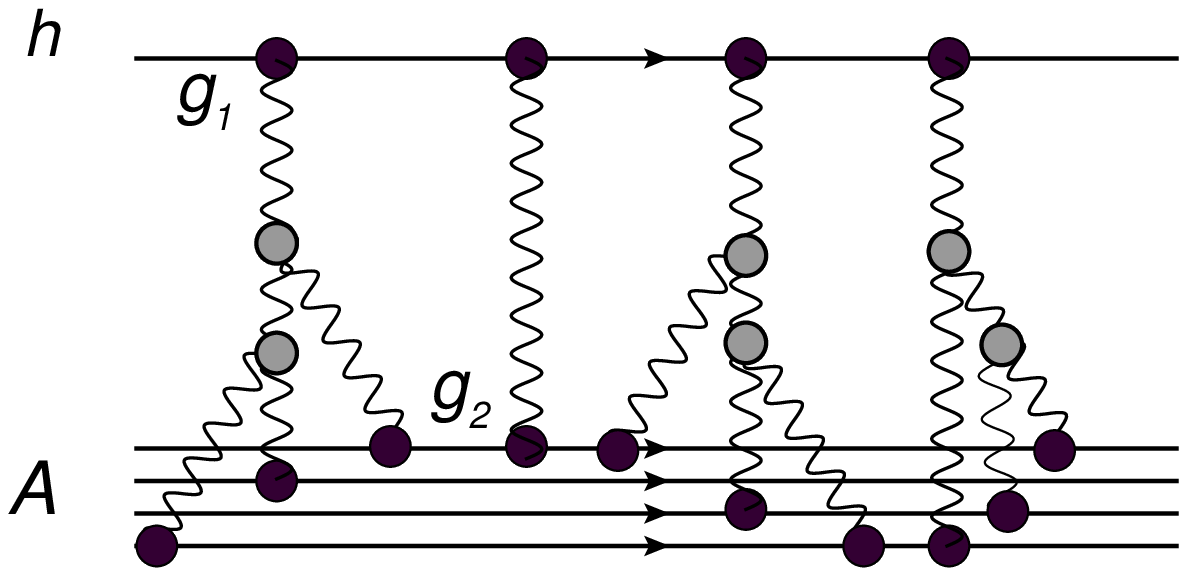,width=110mm}}
\caption{The  set of diagrams that contribute to the scattering amplitude of 
hadron-nucleus scattering in the kinematic region given by \eq{KRHA}.}
\label{hAset}
}
%%%%%%%%%%%%%%%%%%%%%%%%%%%%%%%%%%%%%%%%%%%%%%%%%%%%
For hadron-nucleus collisions we have only one kinematic region 
(\eq{KRHA}) in which we calculate the scattering amplitude. 
This region is similar to the second nucleus-nucleus scattering 
kinematic region (see \eq{KR2}).
\bea \label{KRHA}
&&g_{i}\,S_{A_i}(b)\,G_{3\pom}\,e^{\Delta_{\pom} Y} \,\,\propto\,\,g_i\,
G_{3\pom} A^{1/3}_i\,e^{\Delta_{\pom} Y} \,\,\approx \,\,1;\nonumber\\
&&\,\,\,\,\,\,\,\,\,\,\,\,\,\,\,\,\,\,\,\,\,\,\,\,\,\,\,\,\,\,\,\,
G^2_{3\pom}\,e^{\Delta_{\pom} Y} \,\,\ll\,\,1.
\eea
In this kinematic region, the hadron-nucleus 
scattering amplitude can be written in an eikonal form in which the 
opacity $\Omega$ is given by sum of 
the 'fan' diagrams\cite{SCHW} (see \fig{hAset}). 
\beq \label{HA1}
A_{\mbox{hA}}\Lb Y, b\Rb\,\,=
\,\,i \Lb 1\,\,-\,\,\exp \Lb - \frac{\Omega_{\mbox{hA}}\Lb Y; b\Rb}{2}\Rb \Rb, 
\eeq
with
\beq \label{HA2}
\Omega_{\mbox{hA}}\Lb Y; b\Rb \,\,=\,\,\frac{\tilde{q}_h\,\tilde{g} 
G_{enh}(y)\,S_A\Lb \vec{b}\Rb}{1 \,+\,\tilde{g} G_{enh}(y)\,S_A\Lb \vec{b}\Rb}. 
\eeq
\par
Using \eq{HA1} and \eq{HA2}, we obtain that
\bea \label{HA3}
\sigma^{hA}_{tot}\,&=& 2 \int d^2 b\Lb 1\,\,-\,\,\exp \Lb - 
\frac{\Omega_{\mbox{hA}}\Lb Y; b\Rb}{2}\Rb \Rb; \nonumber\\
\sigma^{hA}_{el}\,&=&  \int d^2 b\Lb 1\,\,-\,\,\exp \Lb - 
\frac{\Omega_{\mbox{hA}}\Lb Y; b\Rb}{2}\Rb \Rb^2; \nonumber\\
\sigma^{hA}_{in}\,&=&  \int d^2 b\Lb 1\,\,-\,\,\exp \Lb - 
\Omega_{\mbox{hA}}\Lb Y; b\Rb\Rb \Rb.
\eea
The processes of diffractive production have been discussed in 
Refs.\cite{BGLM,BORY}.
%%%%%%%%%%%%%%%%%%%%%%%%%%%%%%%%%%
\section{Main formulae}
%%%%%%%%%%%%%%%%%%%%%%%%%%%%%%%%%%
\par
In this paper \eq{MPSI91} replaces the Glauber-Gribov eikonal formula 
to describe the experimental data. However,
we need to adjust this formula to our description of  
hadron-hadron data given in Ref.\cite{GLMM}.
In this paper we use two ingredients that 
were not taken into account in \eq{MPSI91}:
\newline
1) A two channel Good-Walker model\cite{GW} which is 
exclusively responsible for low mass diffraction.
\newline
2) Enhanced Pomeron diagrams that lead to 
a different Pomeron Green's function. This mechanism is 
the main contributor to high mass diffraction.
\newline
\par
In the two channel model we assume that the observed physical 
hadronic and diffractive states are written in the form 
\beq \label{MF1}
\psi_h\,\,=\,\,\alpha\,\Psi_1+\beta\,\Psi_2\,;\,\,\,\,\,\,\,\,\,\,
\psi_D\,\,=\,\,-\beta\,\Psi_1+\alpha \,\Psi_2, 
\eeq
where $\alpha^2+\beta^2\,=\,1$. Note that Good-Walker diffraction 
is presented by a single wave function $\psi_D$.
In our initial approach in which the 
Pomeron interaction with a nucleus proceeds 
through an elastic scattering with a single nucleon, 
we need to replace $\tilde{g}_i$ by 
$\alpha^2 \tilde{g}_i^{(1)} \,+\,\beta^2 \tilde{g}^{(2)}_i$. 
$\tilde{g}_i^{(k)}$  which denotes the vertex of the Pomeron  
interaction with nucleus 1 of the states that have been described 
by either the wave functions $\Psi_1$ or $\Psi_2$.
Since $g_1 = g_2$ we can simplify \eq{MPSI91} 
replacing $\tilde{g}_1$ and $\tilde{g}_2$ by 
$\tilde{g}_1 = \tilde{g}_2 = \tilde{g} 
= \alpha^2 \tilde{g}^{(1)} \, + \, \beta^2 \,\tilde{g}^{(2)}$.
\par
In the framework of our approach we consider 
$G^2_{3\pom}\,\exp\Lb \Delta_{\pom} Y\Rb \,\,\ll\,\,1$ and, 
therefore, we can use the Pomeron's Green 
function written in the form
\beq \label{MF11}
G \left(Y\right)\,\,\,=\,\,e^{\Delta_{\pom} Y}. 
\eeq
In Ref.\cite{GLMM} we sum all enhanced Pomeron diagrams.  
This leads to the replacement
of  the 'bare' Pomeron Green function, 
$G\Lb Y\Rb\,=\,\exp\Lb \Delta_{\pom} Y\Rb$,
by the Green function that sums the enhanced diagrams  
\beq \label{MF2}
\gamma\, G \Lb Y\Rb\,\,\longrightarrow \,\,G_{enh} 
\left(Y\right)\,\,\,=\,\,\,1\,\,-\,\,\exp \Lb \frac{1}{T(Y)}\Rb\,
\frac{1}{T(Y)}\,\,\Gamma\Lb 0,\frac{1}{T(Y)} \Rb.
\eeq
$\Gamma\Lb 0, x\Rb$ is the incomplete Gamma function 
(see {\bf 8.350 - 8.359} in Ref.\cite{RY})) 
and $T\Lb Y\Rb$ is given by \eq{T}.
\par
 Finally, we have  the main formulae for the total and inelastic 
nucleus-nucleus cross sections
\bea
\sigma_{tot}\Lb A_1+A_2;Y\Rb&=& 2\int d^2 b 
\Lb 1-\exp\left\{-\h\int d^2 b'
\frac{\Lb \tilde{g} S_{A_1}\Lb\vec{b}'\Rb
\tilde{g}S_{A_2}\Lb\vec{b}-\vec{b}'\Rb G_{enh}(Y)\Rb}
{1+G_{enh}(Y)\left[\tilde{g}S_{A_1}\Lb\vec{b}'\Rb+
\tilde{g}S_{A_2}\Lb\vec{b}-\vec{b}'\Rb\right]}\right\}\Rb;
\label{MF3}
\\
\sigma_{in}\Lb A_1+A_2;Y\Rb&=&\int d^2 b
\Lb 1-\exp\left\{-\int d^2 b'
\frac{\Lb \tilde{g} S_{A_1}\Lb\vec{b}'\Rb
\tilde{g}S_{A_2}\Lb\vec{b}-\vec{b}'\Rb\,G_{enh}(Y)\Rb}
{1+G_{enh}(Y)\tilde{g}\left[S_{A_1}\Lb\vec{b}'\Rb+
S_{A_2}\Lb\vec{b}-\vec{b}'\Rb\right]}\right\}\Rb.
\label{MF4}
\eea
with $\tilde{g}=\alpha^2\tilde{g}^{(1)}+\beta^2 \tilde{g}^{(2)}$.
%%%%%%%%%%%%%%%%%%%%%%%%%%%%%%%%%%%%%%%%%%%%%%%%%%%%%%%%%%

\FIGURE[h]{\begin{minipage}{50mm}{
\centerline{\epsfig{file=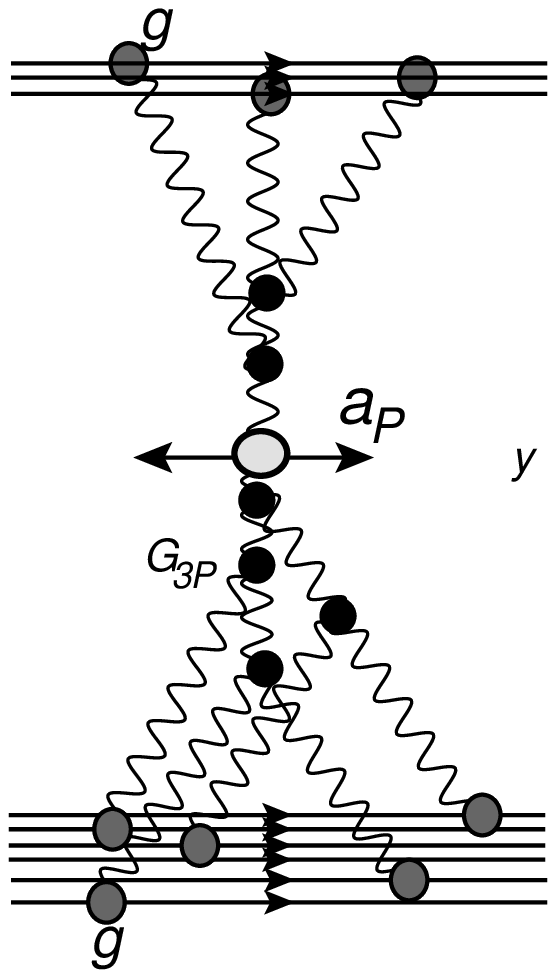,width=30mm}}
\caption{The  set of the diagrams that contribute to the inclusive production of hadrom in ion-ion collisions.}
\label{incl}}
\end{minipage}
}

%%%%%%%%%%%%%%%%%%%%%%%%%%%%%%%%%%%%%%%%%%%%%%%%%%%%%%%%%%%
\par
Using the AGK cutting rules \cite{AGK}, we obtain the formula 
for inclusive production (see the Mueller diagrams\cite{MUDI} 
for the process shown in \fig{incl}). 
The general formula for the inclusive cross section 
has the form 
\beq \label{MF5}
\frac{1}{\sigma_{in}(Y)}\,\frac{d \sigma}{d y} \,\,=
\,\,a_\pom\,\, \frac{\sigma^{MFA}_{in}\Lb Y - y;A_1\Rb\,\sigma^{MFA}_{in}
\Lb y; A_2\Rb}{\sigma_{in}(Y)}.
\eeq
Using \eq{MPSI1} and \eq{MPSI2} one obtains  
\beq \label{MF6}
\sigma^{MFA}_{in}\Lb y; A\Rb\,\,=
\,\,\int d^2 b \frac{\tilde{g} G_{enh}(y)\,S_A\Lb \vec{b}\Rb}
{1 \,+\,\,\tilde{g} G_{enh}(y)\,S_A\Lb \vec{b}\Rb}.
\eeq

\newpage
%%%%%%%%%%%%%%%%%%%%%%%%%%%%%%%%%%%%%%%%%%%%%%%%%%%%%%%%%%%%
\section{Comparison with the experimental data and predictions}
%%%%%%%%%%%%%%%%%%%%%%%%%%%%%%%%%%%%%%%%%%%%%%%%%%%%%%%%%%%
%%%%%%%%%%%%%%%%%%%%%%%%%%%%%%%%%%%%%%%%%%%%%%%%%%%%%%%%%%%%%%
\subsection{Nucleus-nucleus collisions}
%%%%%%%%%%%%%%%%%%%%%%%%%%%%%%%%%%%%%%%%%%%%%%%%%%%%%%%%%%%%%%
The new results presented in this paper are given in  
\eq{MF3} and \eq{MF4}. As noted they can be 
re-written in a Glauber-like form, 
\bea
\sigma_{tot}\Lb A_1+A_2;Y\Rb&=&2\int d^2b 
\Lb 1-\exp\left\{-\h \int d^2b'
\frac{\Lb \sigma^{NN}_{tot}S_{A_1}
\Lb \vec{b}'\Rb S_{A_2}\Lb \vec{b}-\vec{b}'\Rb\,G_{enh}(Y)\Rb}
{1+G_{enh}(Y) \left[\tilde{g}\,S_{A_1}\Lb\vec{b}'\Rb+\tilde{g}S_{A_2}
\Lb\vec{b}-\vec{b}'\Rb\right]}\right\}\Rb; 
\label{CEP1}
\\
\sigma_{in}\Lb A_1+A_2; Y\Rb&=&\int d^2b
\Lb 1-\exp\left\{-\int d^2b'
\frac{\Lb \sigma^{NN}_{in}S_{A_1}\Lb\vec{b}'\Rb S_{A_2}
\Lb\vec{b}-\vec{b}'\Rb\,G_{enh}(Y)\Rb}
{1+G_{enh}(Y)\tilde{g}\left[S_{A_1}\Lb\vec{b}'\Rb+S_{A_2}
\Lb\vec{b}-\vec{b}'\Rb\right]}\right\}\Rb.
\label{CEP2}
\eea
%%%%%%%%%%%%%%%%%%%%%%%%%%%%%%%%%%%%%%%%%%%%%%%%%%%%%%%%%%%%%%%%%%
\FIGURE[h]{\begin{minipage}{50mm}{
\centerline{\epsfig{file=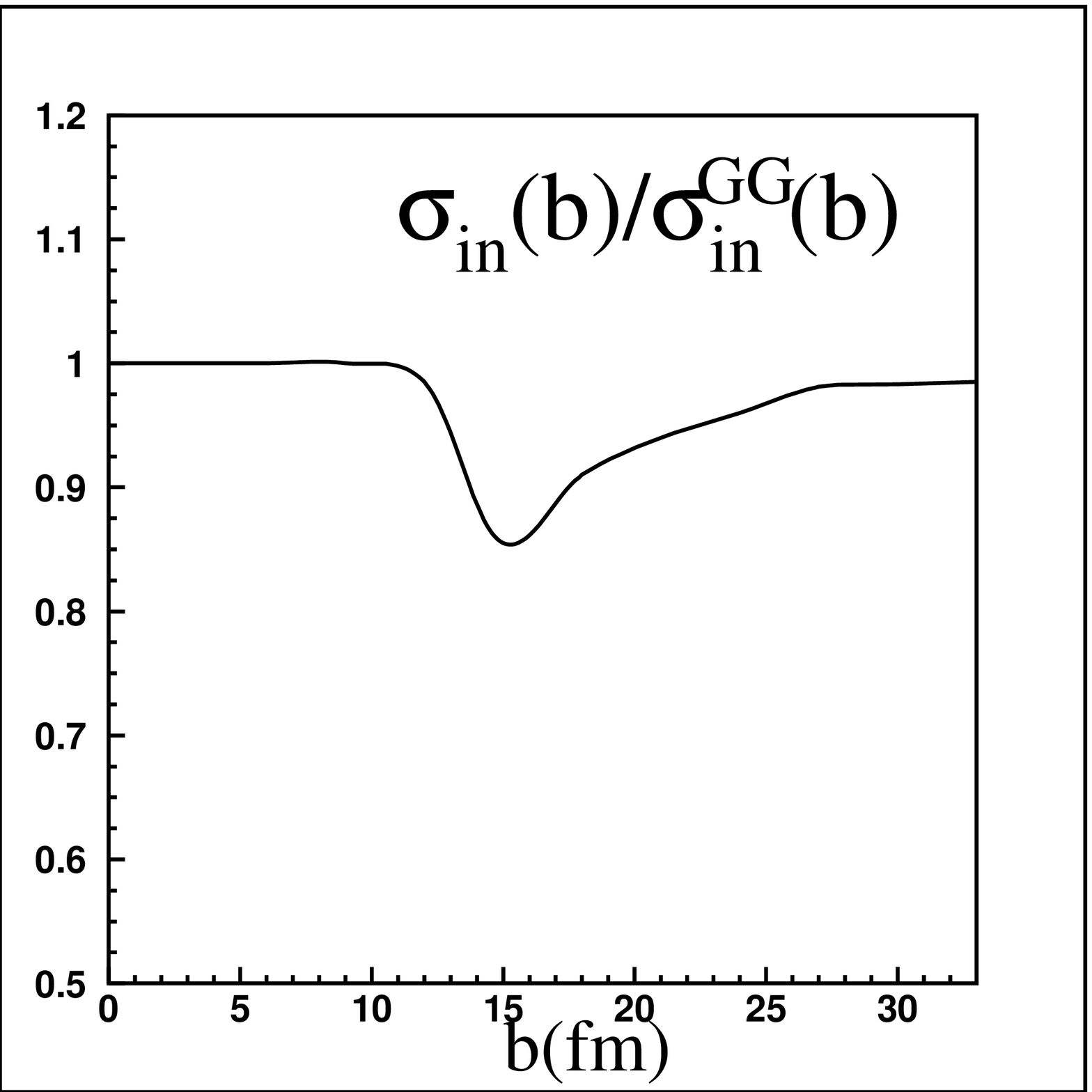,width=50mm}}
\caption{The ratio between the inelastic Au-Au cross sections, 
calculated from the exact formula, the Glauber-Gribov eikonal approach. 
as a function of the impact parameter $(b)$.} 
\label{ratst}}
\end{minipage}
}
%%%%%%%%%%%%%%%%%%%%%%%%%%%%%%%%%%%%%%%%%%%%%%%%%%%%%%%%%%%%%%%%%%
$\sigma^{NN}_{tot}$ and $\sigma^{NN}_{in}$ 
are the total and inelastic nucleon-nucleon cross sections. 
Note that the two sets of nucleus-nucleus cross section 
calculations are only mildly different.
A slightly larger local difference is observed  
in a restricted region of higher impact parameters. 
Indeed, for $b\, \ll\, R_A$ the partial amplitude for nucleus-nucleus 
scattering in either the Glauber-Gribov approach or 
in \eq{CEP1} and \eq{CEP2}, is very close to 1. 
At large values of $b$, \eq{CEP1} and \eq{CEP2}, as well as 
the corresponding Glauber-Gribov cross sections, approach the same limit. 
A small difference between the two approaches 
we have studied may, be detected only at $b \approx 2\,R_A$ 
(see \fig{ratst}). Note, in addition, that the contribution to 
the inelastic cross section in this region of $b$ is very small 
(about 2\% at RHIC energies).
\par
Our results suggest that sensitive observables in  
nucleus-nucleus collision are the number of participants 
and the nuclear modification factor (NMF).
It is well known, that the number of participants $N_{part}$ 
for $A_1-A_2$ scattering is equal to
\beq \label{CEP3}
N_{part}(b) \,=\,\frac{A_1 \,\sigma_{in}^{A_2}\Lb b\Rb\,\,+
\,\,A_2 \,\sigma_{in}^{A_1}\Lb b\Rb}{\sigma_{in}^{A_1 A_2}\Lb b \Rb},
\eeq
where $\sigma_{in}^{A_i}$ are given by \eq{HA3} and $\sigma_{in}^{A_1 A_2}
\Lb b \Rb$ by \eq{CEP2}.  
For Au-Au collisions at RHIC energies (W= 200 \,GeV), the $b=0$ ratio  
$\frac{N_{part}(b)}{N^{GG}_{part}(b)}\,=\,0.93$ with  
$g \,G_{3\pom}/\Delta_{\pom}$ obtained from Ref.\cite{GLMM}. 
In this estimate we take 
$\sigma^{NN}_{in}=\sigma^{NN}_{tot}-\sigma^{NN}_{el}=42\,mb$ 
and for $S_A(b)$ we use the Wood-Saxon parameterization\cite{WS}, 
\beq \label{CEP4}
S_A(b)\,\,=\,\,\int^{+ \infty}_{- \infty}\,d z \frac{\rho_0}{1\,\, +
\,\,e^{\frac{\sqrt{z^2 + b^2} - R_A}{h}}}, 
\eeq
with $\rho_0 = 0.171\,1/fm^3$, $R_A = 6.39\,fm$ and $h = 0.53\,fm$ for Au.  
In this calculation we took 
$\sigma^{NN}_{in}=\sigma^{NN}_{tot}-\sigma^{NN}_{el}-\sigma^{NN}_{diff}=36\,mb$. 
Note that, $\sigma^{NN}_{diff}\,=\,2\sigma^{NN}_{sd}\,+\,\sigma^{NN}_{dd}.$ 
The above estimate is more reasonable than the qualitative estimate of 
$42\,mb$ quoted earlier (see Ref.\cite{KOP}). 
$N^{GG}_{coll}$ has been taken from Ref.\cite{KHNA}. 
We obtain $N_{part}(b=0)/N^{GG}_{part}(b)\,=\,0.90$.
\par
The NMF is defined as
\beq \label{NMF}
R_{AA}\,=\,\frac{\frac{d^2 \sigma_{A_1A_2}}{d y d^2 p_\perp}}
{\frac{d^2 \sigma_{NN}}{d y d^2 p_\perp}}\,=\,\frac{1}{N_{coll}}\,
\,\frac{\frac{d^2 N_{A_1A_2}}{d y d^2 p_\perp}}{\frac{d^2 N_{NN}}
{d y d^2 p_\perp}},
\eeq
where $N$ is the hadron multiplicity. 
In the last equation both $\frac{d^2 N_{A_1A_2}}{d y d^2 p_\perp}$ and 
$\frac{d^2 N_{NN}}{d y d^2 p_\perp}$ are measured experimentally, 
while the number of collisions defined as 
$N_{coll} = A \sigma^{NN}_{in}/\sigma^{A_1 A_2}_{in}$ 
has to be calculated. 
As we have observed for the case of a nucleus-nucleus collision,
the Glauber-Gribov approach to $N_{coll}$ gives the same result as 
the correct formulae, \eq{CEP1} and \eq{CEP2}.
\par
Using \eq{MF5} and \eq{MF6} we calculate the value of $R_{AA}$,
\beq \label{NMF1}
R_{AA}\,=\,\Lb \int d^2 b\,\frac{S_{A_1}\Lb b\Rb}
{1\,+\,\tilde{g}\,G_{enh}\Lb Y/2 - y\Rb\,S_{A_1}\Lb b\Rb}\Rb
\,\Lb \int d^2 b\,\frac{S_{A_2}\Lb b\Rb}{1\,+\,\tilde{g}\,G_{enh}
\Lb Y/2\,+\,y\Rb\,S_{A_2}\Lb b\Rb}\Rb.
\eeq
\par
In Table 1 we display the estimates of $R_{AA}$ 
for Au-Au collisions, calculated in the approach of Ref.\cite{GLMM}. 
In this approach the contribution of the triple Pomeron interactions 
are relatively small, leading to a $1-2\, mb$ contribution,  
which is just a fraction of the calculated   
single inclusive diffractive cross section.
%%%%%%%%%%%%%%%%%%%%%%%%%%%%%%%%%%%%%%%%%%%%
\TABLE{
\begin{tabular}{|l | l l l l l l |}
\hline
\hline
$\gamma$\slash y &0 &0&1 & 2 & 3 & 4\\
 &inclusive & centrality $0 -10 \%$ & & & & \\
\hline
$\gamma_0$ & 0.61 &0.59 &  0.60 & 0.57 & 0.52 & 0.46 \\
\hline
2 $\gamma_0$ & 0.42 &0.39& 0.41 & 0.385 &0.345 & 0.30\\
\hline
3 $\gamma_0$ & 0.31 &0.28& 0.305 &0.28 & 0.25 &  0.218\\
\hline \hline
\end{tabular}
\caption{Inclusive $R_{AA}$ for Au-Au collisions at $W = 200\,GeV.$}
}
%%%%%%%%%%%%%%%%%%%%%%%%%%%%%%%%%%%%%%%%%%%%%%%%%%%%%%%%%
In the above,
$\tilde{g} G_{3\pom}/\Delta_{\pom} = n\tilde{g}\gamma$ (n=1,2,3)
and $\gamma=\gamma_0$ is the value obtained
from the data analysis of the proton-proton scattering in Ref.\cite{GLMM}.
$Y = \ln (s/s_0)$ and $s = W^2$, where $W$ denotes the center mass 
energy. Recall that $R_{AA}$ does not depend on $p_\perp$. 
However, we can trust our approach only at small values of $p_\perp$. 
\par
In Table 1 we checked the single diffraction cross sections
initiated by  triple Pomeron interactions corresponding to 
$\gamma$ values ranging from $\gamma_0$, obtained in Ref.\cite{GLMM}, 
to $3\gamma_0$. Fitting the data with $\gamma$ fixed at either of these 
values we conclude that 
the variance of the overall $\chi^2/d.o.f.$ 
we have obtained is not large. i.e. the data is not very sensitive to 
the value of $\gamma$ within the above range. 
We believe, therefore, that the estimates 
presented in the Table 1 are instructive for obtaining an approximate 
value of $R_{A,A}$. 
For NMF at fixed centrality we have used Ref.\cite{KHNA} relations between the 
centrality cuts and the essential impact parameter region. Note that 
the corrections to $N_{coll}$ using the correct formula are small. 
One can see from Table 1 that $R_{AA}$, in the centrality region 
$(0 - 10\%)$, is only slightly suppressed  in comparison
to the inclusive NMF (see also \eq{NMF}). 
\par
In \fig{rAA1} we plot the data for inclusive $R_{AA}$
(see Refs.\cite{BRAHMS0,BRAHMS1,BRAHMS2,PHENIX1}) 
as compared with our predictions for $W=200\,GeV$.
We checked two values of the triple Pomeron vertex: one which, 
is taken from Ref.\cite{GLMM} ($\gamma=\gamma_0$), and the second 
is 3 times larger. 
There are two different interpretations of the 
physical meaning of our results: 
\newline
1) A traditional one, in which  Pomeron calculus is responsible 
for the structure of the initial partonic wave function of the 
fast hadron, and/or nucleus. Therefore, we need to divide the 
experimental values of $R_{AA}$ by the calculated NMF, 
and explain this ratio $R^{exp}_{AA}/R^{theory}_{AA}$  
by accounting for the final state interactions such as 
jet quenching, energy losses and so on\cite{ELOSS,BDMS}.\newline 
2) In the interpretation which we follow, the Pomeron calculus 
initiates the Color Glass Condensates\cite{GLR,MUQI,MV} 
in the region of large distances. In this case, it gives the correct 
normalization of$R_{AA}$ in the region of small $p_\perp \ll Q_s$.  
$Q_s$ is the saturation momentum. 
The ratio $R^{exp}_{AA}/R^{theory}_{AA}$ can be interpreted as 
originating from two possible sources: a proper account of 
the transverse momentum structure of the parton densities in the 
saturation region (see Refs.\cite{KHNA,KLN,KKT}), 
and/or the final state interactions \cite{ELOSS,BDMS}. 
%%%%%%%%%%%%%%%%%%%%%%%%%%%%%%%%%%%%%%%%%%%%%%%%%%%%%%%%%%%%%%%%
\DOUBLEFIGURE[ht]{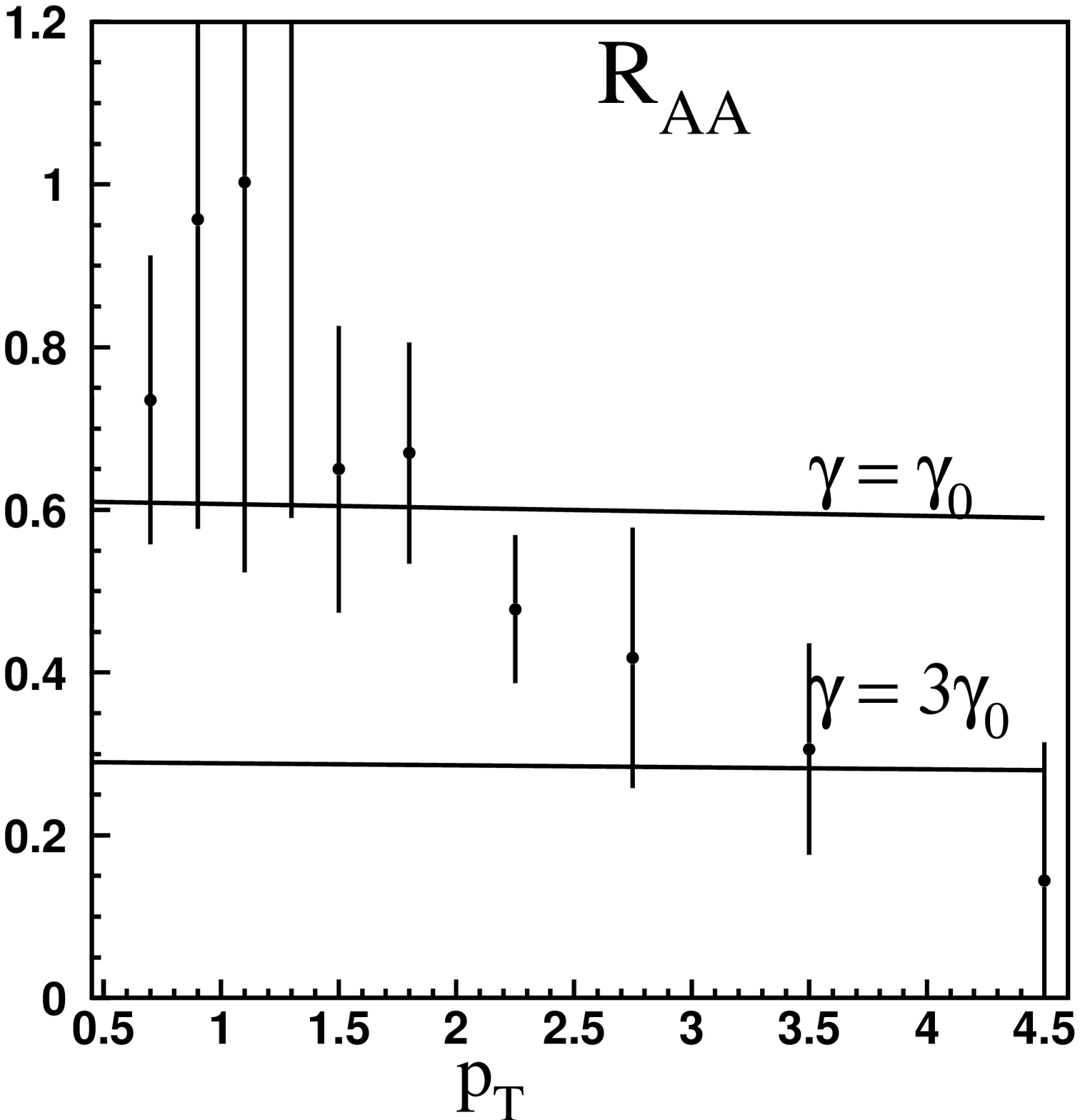,width=80mm,height=60mm}
{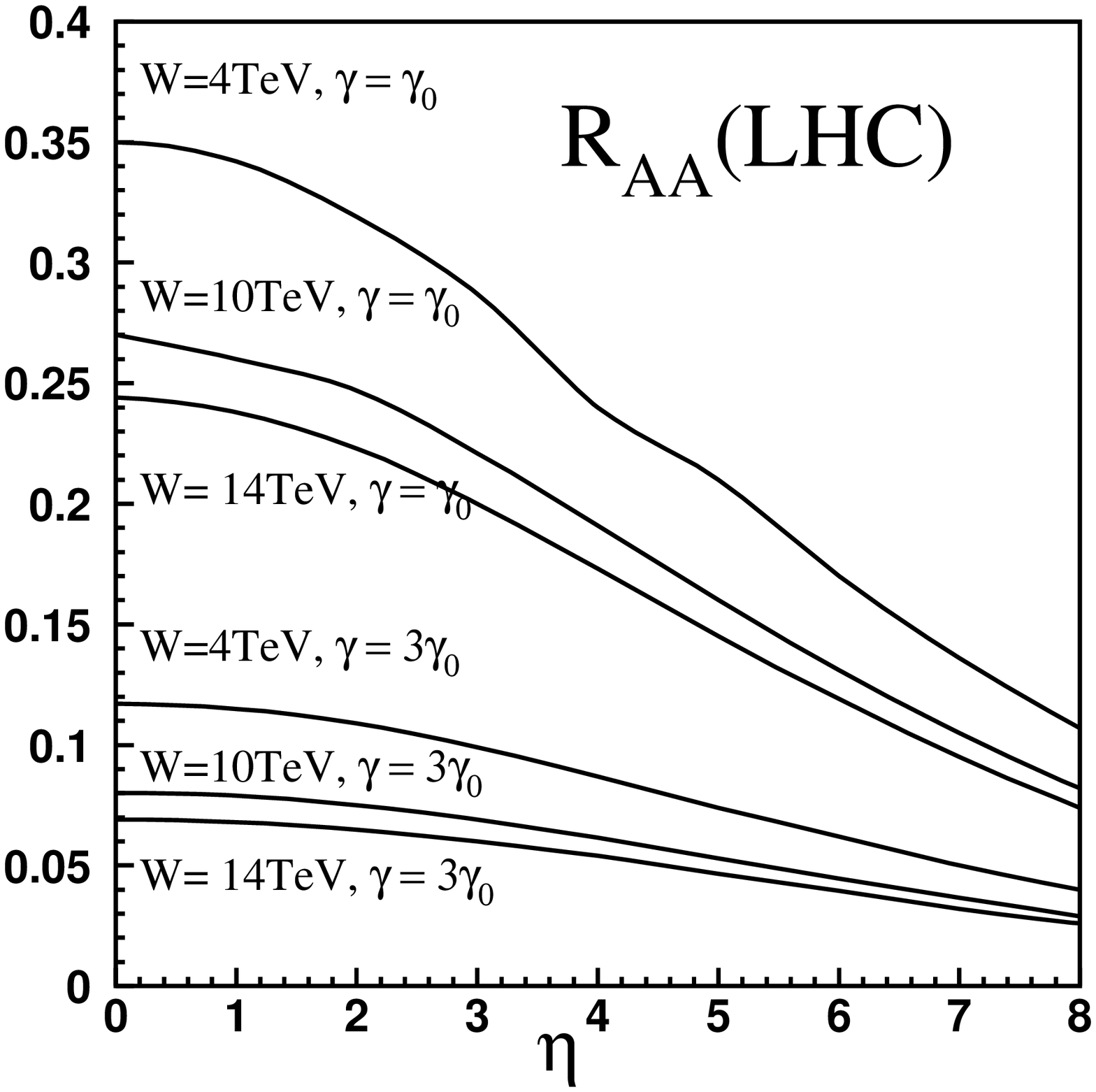,width=80mm,height=60mm}{The experimental value of the $R_{AA}$ 
(the data are taken from Ref.\protect\cite{PHENIX1}) at $W = 200\,GeV$. 
The horizontal lines correspond to two different values 
of the triple Pomeron vertex. 
The value of $\gamma_0$ as well as other parameters such as $\tilde{g}$ 
is taken from Ref.\protect\cite{GLMM}.\label{rAA1}}
{The prediction for the inclusive $R_{AA}$ at the LHC energies. 
The value of $\gamma_0$ as well as other parameters such as $\tilde{g}$ 
is taken from Ref.\protect\cite{GLMM}.\label{rAA2}}.
%%%%%%%%%%%%%%%%%%%%%%%%%%%%%%%%%%%%%%%%%%%%
\subsection{Hadron-nucleus collisions}
%%%%%%%%%%%%%%%%%%%%%%%%%%%%%%%%%%%%%%%%%%
The NMF for proton-nucleus collision is defined by the same 
expression as for nucleus-nucleus, 
\beq \label{NMF3}
R_{pA}\,=\,\frac{\frac{d^2 \sigma_{p A}}{d y d^2 p_\perp}}
{\frac{d^2 \sigma_{pp}}{d y d^2 p_\perp}}\,=\,\frac{1}{N_{coll}}
\,\frac{\frac{d^2 N_{p A }}{d y d^2 p_\perp}}{\frac{d^2 N_{pp}}
{d y d^2 p_\perp}}.
\eeq
This ratio can be  calculated from \eq{MF6} which lead to the following equation
\beq \label{NMF4}
R_{pA}\Lb Y,y\Rb\,=\,\frac{\frac{d^2 \sigma_{p A}}
{d y d^2 p_\perp}}{\frac{d^2 \sigma_{pp}}{d y d^2 p_\perp}}\,=\,\int d^2 b 
\frac{\tilde{g} G_{enh}(y)\,S_A\Lb \vec{b}\Rb}
{1\,+\,\tilde{g} G_{enh}(y)\,S_A\Lb \vec{b}\Rb},
\eeq
where $Y = \ln(s/s_0)$ and $y$ denotes the rapidity of a produced particle. 
However, $N_{coll}$ for 
proton-nucleus scattering turns out to be different from the 
Glauber-Gribov approach estimates that have been used by the experimentalists. 
It means that we have to calculate 
\beq
R^{eff}_{pA}\Lb Y\Rb \,\equiv \,R_{pA}\Lb Y,y\Rb\,N^{theory}_{pA}/N^{GG}_{pA}, 
\eeq
where
\beq \label{MNF4}
N^{theory}_{pA}\Lb Y\Rb\,=\,A\,\frac{\sigma_{in}^{pp}}
{\int d^2 b\,\,\frac{\tilde{g} G_{enh}(Y)\,S_A\Lb \vec{b}\Rb}
{1\,+\,\tilde{g} G_{enh}(Y)\,S_A\Lb \vec{b}\Rb}}.
\eeq
For RHIC energy range we can replace $\,N^{theory}_{pA}/N^{GG}_{pA}$ by 
$\sigma^{PP}_{in}(\mbox{theory})/\sigma^{PP}_{in}$. 
As we saw, a common approximation\cite{KHNA},
$\sigma^{pp}_{in}(GG)=\sigma_{tot}-\sigma_{el}=42\,mb$, 
in Glauber-Gribov approach calculations is improved by\cite{KOP}   
$\sigma^{pp}_{in}(theory)=\sigma_{tot}-\sigma_{el}-
\sigma_{diff}(GW)\,=36\,mb$, which is 
suitable for the approach of Ref.\cite{GLMM}. 
Recall that $\sigma_{diff}\,=\,2\sigma_{sd}\,+\,\sigma_{DD}$. 
GW denotes a contribution initiated by the Good-Walker mechanism, 
see Ref.\cite{GLMM} for details and references.
This difference induces a correcting coefficient 
$\,N^{theory}_{pA}/N^{GG}_{pA} = 36/42 = 0.86$.
However, in the framework of our approach we need to  
 modify the total and quasi-elastic 
cross sections due to the interaction with a nucleus 
rather than a nucleon (see \fig{modn}). 
The correct inclusion of this interaction leads to
\bea \label{MNF5}
&&\sigma_{in}\Lb p+A; Y\Rb\,\,=\\
&&\int d^2 b \Lb 1 \,-\,\exp \Lb - 
\left\{ \sigma^{pp}_{tot} \frac{S_A(b)}{\Lb 1 \,+\,\tilde{g}\,G_{enh}
\Lb Y\Rb \,S_A(b)\Rb} \,-\,(\sigma^{pp}_{el}\,+\,\sigma^{pp}_{diff}) 
\frac{S_A(b)}{\Lb 1 \,+\,\tilde{g}\,G_{enh}\Lb Y\Rb \,S_A(b\Rb)^2}\right\} \Rb 
\Rb. \nonumber
\eea
%%%%%%%%%%%%%%%%%%%%%%%%%%%%%%%%%%%%%%%%%%%%%%%%%%%%%%%%
\FIGURE[h]{\begin{minipage}{80mm}{
\centerline{\epsfig{file=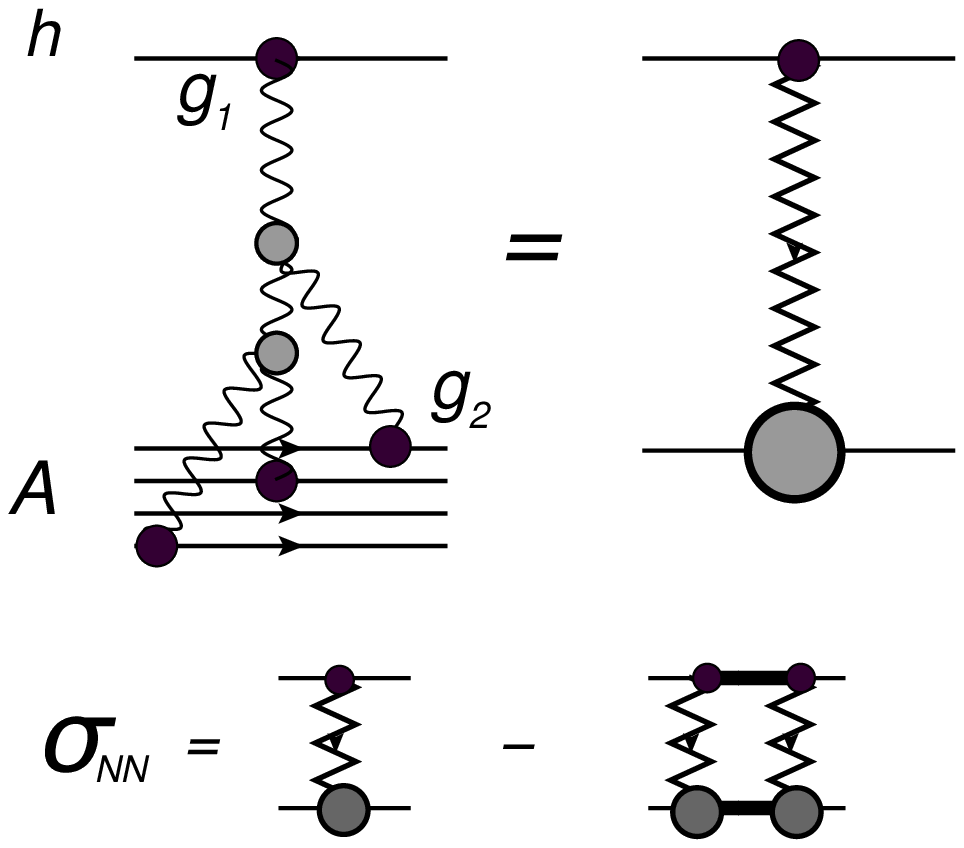,width=70mm}}
\caption{ The modified nucleon-nucleon inelastic cross section in our approach.}
\label{modn}}
\end{minipage}
}
%%%%%%%%%%%%%%%%%%%%%%%%%%%%%%%%%%%%%%%%%%%%%%%%%%%%%%%%
%%%%%%%%%%%%%%%%%%%%%%%%%%%%%%%%%%%%%%%%%%%%%%%%%%%%%%%%%%%%%%%%
\TABLE[ht]{\begin{tabular}{| l |l |l l l l l l|}
\hline \hline
$\gamma$\y & $N^{theory}_{coll}/N^{GG}_{coll}$ & 0 & 1& 2& 3& 4& 5\\
\hline
 $\gamma_0$ & 0.9 & 0.71 & 0.66 & 0.60 & 0.53 & 0.46 & 0.39 \\
 $2 \gamma_0$ & 0.81 & 0.53 & 0.495 & 0.44 & 0.38 & 0.32 & 0.26 \\
$3 \gamma_0$ & 0.73 & 0.415 & 0.36 & 0.30 & 0.25 & 0.20& 0.16 \\
\hline \hline
\end{tabular}
\caption{$N^{theory}_{coll}/N^{GG}_{coll}$ and $R^{eff}_{pA}\Lb Y\Rb\,\equiv\,
R_{pA}\Lb Y,y\Rb\,N^{theory}_{pA}/N^{GG}_{pA}$ as function of 
rapidity for different values of $G_{3\pom}$.}
}
%%%%%%%%%%%%%%%%%%%%%%%%%%%%%%%%%%%%%%%%%%%%%%%%%%%%%%%%%%%%%%%%%%
Table 2 presents our calculations for $N^{theory}_{coll}/N^{GG}_{coll}$. 
%and $R^{eff}_{pA}\Lb Y\Rb$.
As we have discussed, we can only trust these estimates at low transverse 
momenta. 
Comparing $R^{eff}_{pA}\Lb Y\Rb$ 
with the experimental data (see Ref.\cite{BRAHMS1}), 
one can see that we reproduce 
the low transverse momenta data well. 
As in the case of nucleus-nucleus scattering, we believe 
that our estimates 
give the correct normalization for the 
NMF at low $p_\perp$, 
leaving the explanation of the $p_\perp$ 
dependence of $R_{pA}$ to the dependence of the parton 
densities in the saturation region and beyond
(see Ref.\cite{KKT}), and the jet quenching and energy 
loss in the final state\cite{ELOSS,BDMS}.
If we divide all data by our calculated $R^{eff}_{pA}$,
the $p_\perp$ dependence of the NMF resembles the ordinary 
Cronin effect\cite{Cronin} which approaches 
unity at large $p_\perp$.
%%%%%%%%%%%%%%%%%%%%%%%%%%
\section{Conclusions}
%%%%%%%%%%%%%%%%%%%%%%%%%%
The main theoretical results of this paper are the formulae for the 
total and inelastic cross sections of
nucleus-nucleus scattering in the kinematic region of \eq{KR2}.
These formulae together with equations for hadron-nucleus total and inelastic 
cross section and for inclusive production, enable us 
to make comparison with the experimental data at RHIC energies. 
This comparison suggests that the effect of the initial partonic
wave function can explain the essential part of the nucleus modification factor. 
Consequently, jet quenching
and energy losses are responsible for a comparatively
small part of the nuclear suppression. 
Since we developed our approach based on soft Pomeron calculus,
we discuss how our approach is related to other approaches on the market. There
are two alternative points of view on this subject. In the traditional one, 
the soft Pomeron is a separate issue. 
The estimates based on soft Pomeron approach provides the information on 
the partonic wave function
in the initial state. Therefore, in such an approach, 
the difference between the experimental data and
our estimates should be explained by the interactions in the final state. 
We support the second point
of view in which the soft Pomeron approach that has been developed in 
this paper and in Ref.\cite{GLMM}, 
is a natural generalization of the Color Glass Condensate approach, 
and it provides the normalization of
the NMF at long distances. Indeed, in CGC approach the relation between
$N_{part} = c S_\perp Q^2_s$ enters with the coefficient $c$ which could  
only be determined from numerical simulation.
We firmly believe that our approach suggests an alternative 
method to determine the numerics of CGC.
The $p_\perp$ dependence within this interpretation , 
is correlated with the $p_\perp$ dependence of
the partonic densities in the saturation region and beyond. It depends, as well,
on the final state interactions, where we predict only a slight suppression.
\par 
Using our approach we predict the NMF at LHC energies and we hope that this 
prediction will be useful 
for ion-ion interactions at the LHC.

\end{document}